\documentclass{article}
\usepackage{graphicx}
\title{Control software analysis, Part II\\
Closed-loop properties}
\author{Eric Feron~\thanks{Dutton/Ducoffe Professor of Aerospace Software Engineering,
Georgia Institute of Technology. {\tt feron@gatech.edu}} \and Fernando Alegre~\thanks{College of Computing, Georgia Tech. {\tt fernando@cc.gatech.edu}}}

\def\reals{{\bf R}}
\begin{document}
\bibliographystyle{alpha}
\maketitle

{\bf Abstract:} The analysis and proper documentation of the properties of closed-loop control software presents many distinct aspects from the analysis of the same software running open-loop. Issues of physical system representations arise, and it is desired that such representations remain independent from the representations of the control program. For that purpose, a concurrent program representation of the plant and the control processes is proposed, although the closed-loop system is sufficiently serialized to enable a sequential analysis. While dealing with closed-loop system properties, it is also shown by means of examples how special treatment of nonlinearities extends from the analysis of control specifications to code analysis.

\section{Closed-loop software system analysis}
The first part of this report essentially dealt with the analysis of open-loop control systems, that is, the analysis of control algorithms without regards for their stabilizing effect on the plant: In~\cite{Fer1:08}, we were concerned with verifying and quantifying the boundedness of all computed quantities in the controller, regardless of the actual performance of the closed-loop system. In that regard, the analysis is similar to that described in~\cite{CousotEtAl-ESOP05}, although our work is much more focused on control algorithms and, unlike the work in~\cite{CousotEtAl-ESOP05}, it ignores the many other software components present in a fully developed avionics system. In the present document, we focus on the analysis and verification of the system's closed-loop performance, beginning with closed-loop system stability. One of the elements of this report is the modeling, by means of an example, of closed-loop control systems as two concurrent processes, where one process represents the real-time computer program and the other represents the system being controlled. The concurrent representation of these processes does not make their analysis more complicated, and it
provides the engineer with greater flexibility by making the choice of controller representations and implementations independent from the representation of the plant being controlled. Again, the best way to perform the closed-loop system analysis is to begin with specification-level closed-loop system analysis, and then to move on to incorporate more detailed controller implementations. This is the process illustrated in this report.
 
\subsection{Specification-level analyses: Stability}
\begin{figure}[htbp]  
   \hspace{0mm}
   \begin{center}
    \includegraphics[width=10cm]{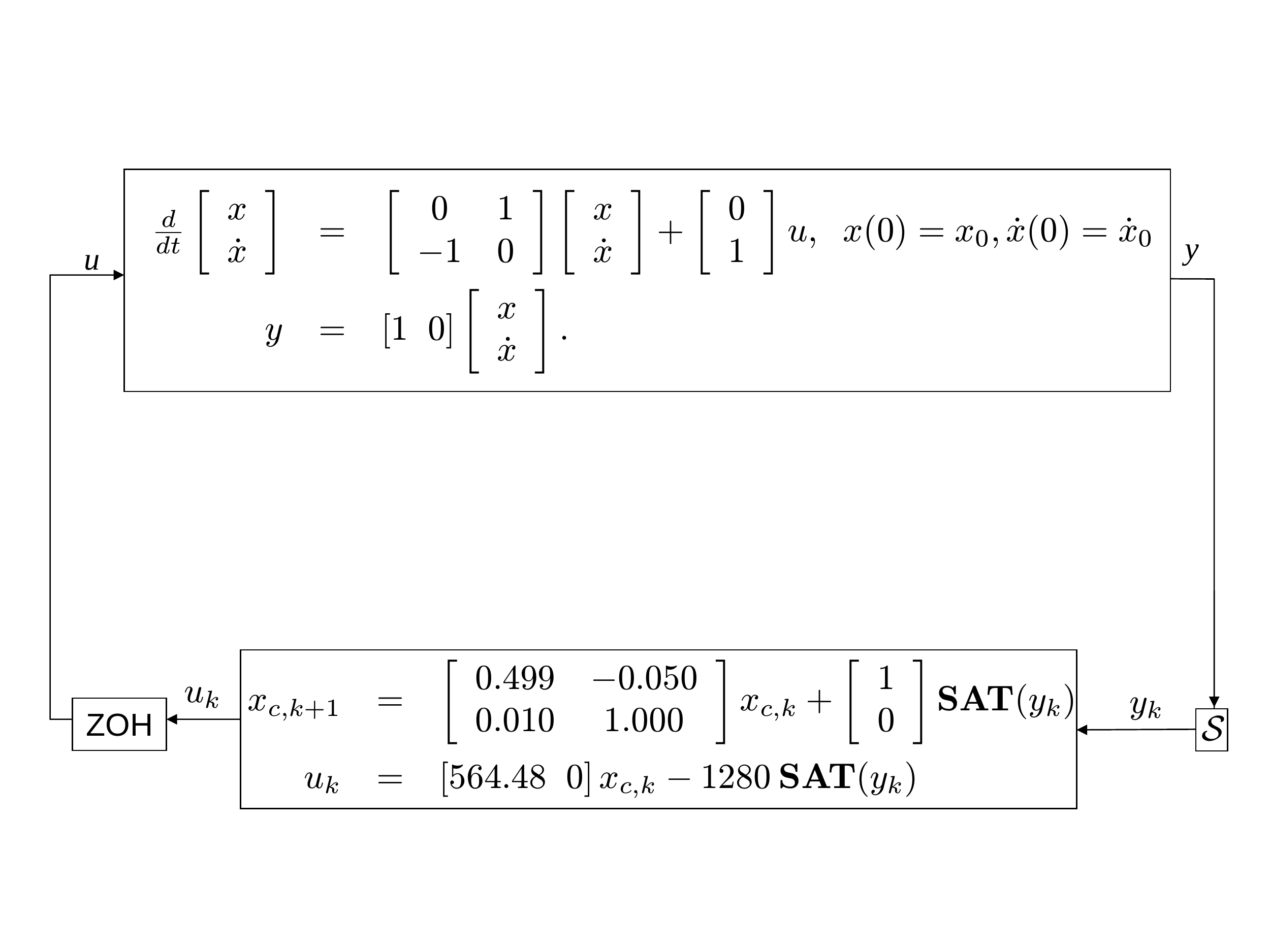}
    \end{center}
    \caption{Feedback system}
    \label{Figure2}
\end{figure}

The analysis of the specifications of the lead-lag controller designed and presented in~\cite{Fer1:08} begins with Figure~\ref{Figure2}. In that figure, the controlled system is represented by a continuous differential equation, while the controller is a discrete-time system that interacts with the continuous system by means of a signal sampler at the input and a zero-order hold at the output. The dynamics of the controller may be written
\[
\begin{array}{rcl}
x_{c,k+1} & = & A_cx_{c,k}+B_c\: {\bf SAT}(y_k)\\
u_k & = & C_c x_{c,k} + D_c \:{\bf SAT}(y_k)
\end{array}
\]
with

\[
\begin{array}{l}
A_c=\left[\begin{array}{cc} 0.4999 & -0.0500 \\ 0.0100 & 1.0000\end{array} \right], \; B_c = \left[\begin{array}{c} 1 \\ 0 \end{array} \right] \\[10pt]
C_c = \left[564.48 \;\; 0\right],\; \mbox{and } D_c= -1280.
\end{array}
\]

\begin{figure}[htbp]  
   \hspace{0mm}
   \begin{center}
    \includegraphics[width=10cm]{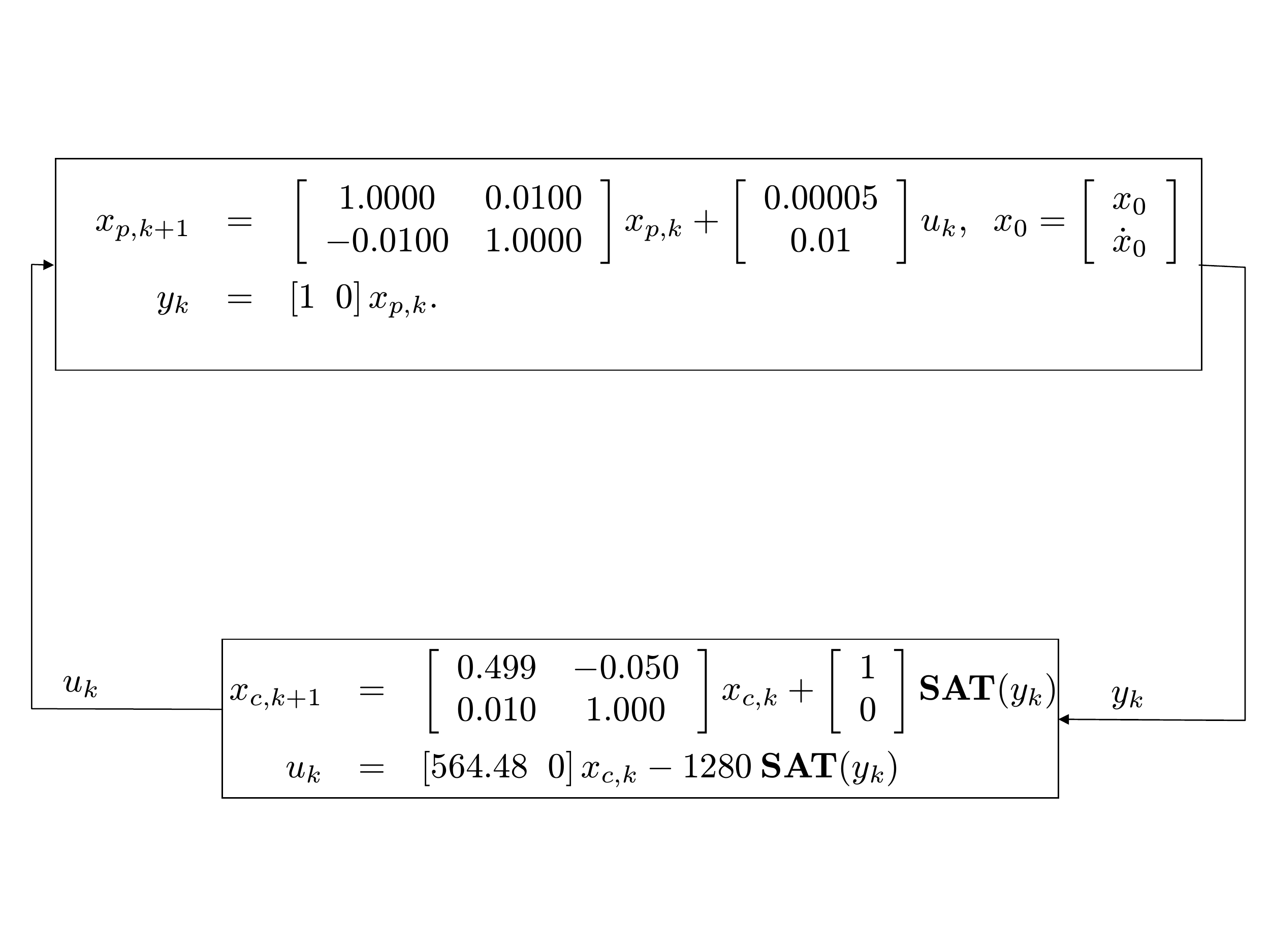}
    \end{center}
    \caption{Equivalent discrete system}
    \label{discrete-system}
\end{figure}

It is well-known from control theory~\cite{FrP:80} that the analysis of such a closed-loop, sampled-data system can be performed by replacing the plant, the zero-order hold and the sampler by the discrete-time system 
\[
\begin{array}{rcl}
x_{p,k+1} &= &A_px_{p,k} + B_pu_{k}\\
y_k &= & C_px_{p,k}
\end{array}
\]
with 
\[
A_p = \left[\begin{array}{cc}
    1.0000  &  0.0100 \\
   -0.0100  &  1.0000 
   \end{array} \right], \; B_p = \left[\begin{array}{c} 0.00005 \\ 0.01\end{array} \right]
\]
and $C_p = \left[1 \;\; 0\right]$. Moreover, $x_{p,k} = x_p(0.01k)$, where $k \in {\bf N}$.
Then the analysis of the system shown in Fig.~\ref{Figure2} is equivalent to the analysis of the discrete-time system shown in Fig.~\ref{discrete-system}.
The stability analysis of this discrete-time system is made more difficult than that of the controller alone~\cite{Fer1:08} because of the presence of a saturation nonlinearity aimed at limiting the range of sensor inputs to the control system. This nonlinearity may be accounted for in the overall system stability analysis by using a variety of computational techniques, such as those described in~\cite{BEFB:94}. In particular one of the available techniques consists of computing a quadratic Lyapunov function for the system. It is, for example, possible to show that the quadratic function
\[
V(x) = \left[\begin{array}{c} x_c \\ x_p \end{array} \right]^TP \left[\begin{array}{c}x_c \\ x_p \end{array} \right], \mbox{ with } P = \left[\begin{array}{cccc}    
     0.2205 &   0.0188  & -0.0750   & 0.0177\\
    0.0188  &  0.4736  &  0.0535  &  0.0015\\
   -0.0750  &  0.0535  &  0.1012  & -0.0049\\
    0.0177  &  0.0015  & -0.0049  &  0.0015
\end{array} \right]
\]
decays along all trajectories $(x_{c,k}, x_{p,k})_{k = 1,2,...}$ for initial conditions where the controller is at rest ($x_{c,0} = 0$) and the initial state $x_{p,0}$ of the mechanical system lies inside the ellipse ${\cal E}_Q$ with 
\begin{equation}
Q = \left[\begin{array}{cc}        0.1012  & -0.0049\\
   -0.0049 &   0.0015 \end{array} \right] \mbox{ and } {\cal E}_Q = \{y \in \reals^2 ~|~y^TQy\leq 1\}
\label{ellips}
\end{equation}
The mechanisms used to reach that conclusion rely on standard stability considerations, most notably absolute stability theory: Consider the closed-loop system dynamics
\[
x_{k+1} = Ax_k + B{\bf SAT}(Cx_k),
\]
with $x = [x_c^T \;\; x_p^T]^T$, 
\[
A = \left[\begin{array}{cc} A_c & 0 \\
B_pC_c & A_p \end{array}\right] = \left[\begin{array}{cccc}      0.4990 &  -0.0500 &        0    &     0\\
    0.0100 &   1.0000  &       0  &       0\\
    0.0282     &    0  &  1.0000  &  0.0100\\
    5.6448     &    0  & -0.0100  &  1.0000
\end{array} \right], 
\]
\[
B = \left[\begin{array}{c} B_c \\ B_p D_c\end{array}\right], \mbox{ and }\; C = [0 \;\; C_p].
\]
Consider the set of $x$'s such that $x^TPx<1$. It is easy to show that (i) the set of all 
$x = [0 \; \; x_p^T]^T$ such that $x_p \in {\cal E}_Q$ is contained in ${\cal E}_P$, and (ii) that for all $x \in {\cal E}_P$, $({\bf SAT} (Cx) -0.2 Cx)( {\bf SAT }(Cx) - Cx)\leq 0$. Introducing $y_{c,k} ={\bf SAT}(Cx_k)$, it is therefore sufficient to show that
\[
x_{k+1}^TPx_{k+1} - x_k^TPx_k\leq 0
\]
whenever $(y_{c,k} -0.2 Cx_k)(y_{c,k} - Cx_k)\leq 0$. An equivalent condition is
\[
\left[\begin{array}{c}x \\ y_c\end{array} \right]^T\left(\left[\begin{array}{c}A^T\\ B^T\end{array}\right]P[A\; B]- \left[\begin{array}{c} I \\ 0 \end{array}\right]P\left[ I \; 0\right]\right)
\left[\begin{array}{c}x \\ y_c\end{array} \right]\leq 0
\]
whenever $(y_c -0.2 Cx)(y_c - Cx)\leq 0$. Using the well-known ${\cal S}$-procedure~\cite{AiG:64}, a sufficient condition is the existence of $\lambda>0$ such that 
\begin{equation}
\left[\begin{array}{c}A^T\\ B^T\end{array}\right]P[A\; B]- \left[\begin{array}{c} I \\ 0 \end{array}\right]P\left[ I \; 0\right]-
\lambda \left[\begin{array}{cc}0.2 C^TC & -0.6 C^T\\ -0.6 C & 1\end{array} \right]
\leq 0.
\label{S-proc}
\end{equation}
The latter inequality can be easily shown to be true for $\lambda = 6.76$. It must be kept in mind that all statements related to numerical objects and computations are made under the assumption that computations on reals are exact. This assumption is obviously false and must be eventually addressed when implementing the operational software verification tools that may arise from this paper, using for example techniques developped by Goubault~\cite{Gou:01}. We are now interested in showing how this proof of (local) closed-loop stability may be translated at the code and system level. To be more precise, we will show how invariance of the ellipsoid ${\cal E}_P$ can be exploited to develop a proof of proper behavior, such as stability and performance, for the computer program that implements the controller, as it interacts with the physical system. Unlike the developments relative to controller open-loop properties, this proof necessarily involves the presence of the controlled artifact. It would not make sense for the system to be represented at various degrees of computer program implementations since these do not have any physical meaning. Thus we choose to represent the plant and the computer program as two concurrent programs, as shown in Table~\ref{programs}.
The computer program representation of the mechanical system will remain invariant, written in a high-level language like MATLAB, while the representation of the controller will be allowed to evolve to reflect the various stages of its implementation, without changing the representation of the physical system. 
We assume the two programs communicate by means of {\tt send / receive} commands. It is assumed that the {\tt receive} command is blocking, that is, the program execution cannot proceed until the variable of interest has been received.
\begin{table}
\vspace{3mm}
{\tt \hspace{-20mm}
\begin{tabular}{l}
\% Controller Dynamics\\
 1c:  Ac = [0.4990, -0.0500; 0.0100,1.0000]; \\
 2c:  Cc=[564.48, 0];\\
 3c:  Bc=[1;0]; D = -1280;  \\
 4c:  xc = zeros(2,1); \\
 5c:	receive(y); \\
 6c:  while (1)  \\
 7c:   $\;\;\;$yc = max(min(y,1),-1);  \\ 
 8c:   $\;\;\;$u = Cc*xc + Dc*yc;  \\
 9c:   $\;\;\;$xc = Ac*xc + Bc*yc; \\
10c:   $\;$send(u);\\
11c:	 receive(y); \\
12c:  end 
\end{tabular}
\begin{tabular}{l}
\% Plant Dynamics \\
 1p:  Ap = [1.0000,0.0100;-0.0100,1.0000];\\
 2p:  Cp=[1,0];\\
 3p:  Bp = [0.00005; 0.01]; \\
 4p:  while (1)\\
 5p:   $\;\;\;$ y = Cp*xp; \\
 6p:   $\;\;\;$ send(y);\\
 7p:	 $\;\;\;$ receive(u);\\
 8p:   $\;\;\;$ xp = Ap*xp + Bp*u;\\
 9p: end;\\
\end{tabular}
}
\caption{Concurrent representation of plant and controller implementations}
\label{programs}
\end{table}
While the programs in Table~\ref{programs} are purely event-driven, it is possible to modify the algorithmic description of the plant to account for the presence of time. The question of establishing proofs of stability of the closed-loop system at the code level 
is necessarily tied to understanding the behavior of the controller and that of the plant. The entire state-space then consists of the direct sum of the controller's state-space and that of the plant. 
We now propose to leverage the approach developed in the first part of this report~\cite{Fer1:08} to document the corresponding system of two concurrent programs. This documentation can be extended to other representations of the control program implementation: Table~\ref{second-table} shows one such implementation of the controller in pseudo-C.
\begin{table}
\begin{verbatim}
/* variable declarations */
double Ac[2][2];
double Cc[2];
double Bc[2];
double Dc;
double xc[2], yc,y,u;
int flags, flaga;

/* aux vars */
int i,j;
double xc_new[2];

/* code */
Ac[0][0]=0.4990; Ac[0][1]=-0.0500; Ac[1][0]=0.0100;Ac[1][1]=1.0000;
Cc[0]=564.48;Cc[1]=0.0;
Bc[0]=1.;Bc[1]=0.;
Dc=-1280.;
xc[0]=0.;xc[1]=0.;
receive(y);
while(1) {
  yc = (y > 1. ? 1. : y) < -1. ? -1. : y;
  u = 0.0;
  for(i=0;i<2;i++) u += Cc[i]*xc[i];
  u += Dc*yc;
  for(i=0;i<2;i++) {
    xc_new[i] = 0.0;
    for(j=0;j<2;j++) xc_new[i] += Ac[i][j]*xc[j];
    xc_new[i] += Bc[i]*yc;
  }
  for(i=0;i<2;i++) xc[i] = xc_new[i];
  send(u);
  receive{y}
}
\end{verbatim}
\caption{Implementation of controller in C}
\label{second-table}
\end{table}
One of the interesting aspects of the processes in this report is their concurrency, which can make the structure of the state transitions rather complicated. However, a close inspection of the programs reveals a relatively simple transition structure. Our investigation thus begins with a forward analysis of both programs. 
Since only local stability has been proven from the program specifications, we may not expect any better result by inspecting and documenting the code. 
Declared but non-initialized variables $v$ will be assumed to belong to the empty set, that is $v \in \perp$. Likewise, the set of all possible values will be denoted $\top$, and 
a variable $v$ that may take any possible value will be characterized as $v \in  \top$.
Since both programs run concurrently, tracking 
state values and the sets they belong to may be somewhat challenging. 
The only variable whose value is assumed to be initialized prior to the plant execution is the initial plant state $x_p$. For the purpose of stability analysis, we assume that the state $x_p$ is initially within the ellipsoid~${\cal E}_Q$, where $Q$ is defined in~(\ref{ellips}).
Thus, prior to the initiation of the plant process, and using insight from the system specification analysis, we form the condition 
$\left\{ (x_c, x_p) \in  {\cal E}_{P}, \;(y,y_c,u)\in \top \right\}$. 
The first three lines of the plant dynamics {\tt 1p, 2p, 3p} do not change this condition, and so they may be commented as

{\tt \vspace{3mm}
\begin{tabular}{l}
$\left\{ (x_c, x_p) \in  {\cal E}_{P}, \;(y,y_c,u)\in \top \right\}$\\
1p:  Ap = [1.0000,0.0100;-0.0100,1.0000];\\
$\left\{ (x_c, x_p) \in  {\cal E}_{P}, \;(y,y_c,u)\in \top \right\}$\\
2p:  Cp=[1,0];\\
$\left\{ (x_c, x_p) \in  {\cal E}_{P}, \;(y,y_c,u)\in \top \right\}$\\
3p:  Bp = [0.00005; 0.01]; \\
$\left\{ (x_c, x_p ) \in  {\cal E}_{P}, \;(y,y_c,u)\in \top \right\}$
\end{tabular}
}\vspace{3mm}

Line {\tt 4p: while (1)}, together with line {\tt 9p: end;} defines the main loop of the plant dynamics. We use the requirements analysis to {postulate} the following set of pre- and post-conditions

\vspace{3mm}
\begin{tabular}{l}
$\left\{ (x_c, x_p) \in  {\cal E}_{P}, \;(y, y_c,u)\in \top \right\}$  \\
{\tt 4p:   while (1)}\\
$\left\{ (x_c, x_p) \in  {\cal E}_{P}, \; (y,y_c,u)\in \top \right\}$  \\
\vdots \\
$\left\{ (x_c, x_p) \in  {\cal E}_{P}, \;(y,y_c,u)\in \top \right\}$  \\
{\tt 9p:  end}\\
$\left\{ false \right\}$
\end{tabular}
\vspace{3mm}

The last assertion $\{ false \}$ simply indicates a never ending loop; this should be expected from a dynamical system that never ceases to execute. In the following, and in order to simplify the notation, we will omit to mention the variables whose value could be arbitrary (that is, the variables whose values belong to $\top$). With this convention, the previous set of commented lines becomes

\vspace{3mm}
\begin{tabular}{l}
$\left\{ (x_c, x_p) \in  {\cal E}_{P} \right\}$  \\
{\tt 4p:   while (1)}\\
$\left\{ (x_c, x_p) \in  {\cal E}_{P} \right\}$  \\
\vdots \\
$\left\{ (x_c, x_p) \in  {\cal E}_{P} \right\}$  \\
{\tt 9p: end}\\
$\left\{ false \right\}$.
\end{tabular}
\vspace{3mm}

The following line, {\tt 5p: y = Cp*xp} assigns a value to the variable $y$. The corresponding pair of conditions is therefore

\vspace{3mm}
\begin{tabular}{l}
$\left\{ (x_c, x_p) \in  {\cal E}_{P} \right\}$  \\
{\tt 5p:  y = Cp*xp}\\
$\left\{(x_c, x_p, y) \in {\cal G}_{R} \right\}$,
\end{tabular}
\vspace{3mm}

with  
\[
{\cal G}_R = \left\{x \in \reals^n \;\left|\; 
\left[\begin{array}{cc} 1 & x^T\\x & R \end{array} \right]\geq 0\right. \right\}.  
\]

The next line, {\tt 6p: send(y);} does not affect the state variables of interest and therefore
becomes

\vspace{3mm}
\begin{tabular}{l}
$\left\{(x_c, x_p, y) \in {\cal G}_{R} \right\}$  \\
{\tt 6p: send(y)}\\
$\left\{(x_c, x_p, y) \in {\cal G}_{R} \right\}$  \\
\end{tabular}
\vspace{3mm}

At this point, the execution of the plant process is blocked by the {\tt receive} command in line {\tt 7p}.
We now turn our attention to the controller code.
The first four lines of the controller code {\tt 1c, 2c, 3c, 4c}, are commented using the statement $\left\{(x_c, x_p, y) \in {\cal G}_R, \;(y_c,u)\in \top \right\}$ or, in short, $\left\{(x_c, x_p, y) \in {\cal G}_R\right\}$. Thus we get

\vspace{3mm}
\begin{tabular}{l}
$\left\{(x_c, x_p,y) \in {\cal G}_R \right\}$  \\
{\tt 1c:  Ac = [0.4990, -0.0500; 0.0100,1.0000];} \\
$\left\{(x_c, x_p,y) \in {\cal G}_R \right\}$  \\
{\tt 2c:  Cc=[564.48  0];}\\
$\left\{(x_c, x_p,y) \in {\cal G}_{R} \right\}$  \\
{\tt 3c:  Bc=[1; 0]; D = -1280;}  \\
$\left\{(x_c, x_p,y) \in {\cal G}_{R} \right\}$,  \\
{\tt 4c:  xc = zeros(2,1);} \\
$\left\{(x_c, x_p,y) \in {\cal G}_{R} \right\}$. 
\end{tabular}
\vspace{3mm}

At this point, the execution of the controller program is halted by the command {\tt 5c: receive(y)} and can proceed only after the plant has executed line {\tt 6p}. Valid pre- and post-conditions are

\vspace{3mm}
\begin{tabular}{l}
$\left\{(x_c, x_p, y) \in {\cal G}_{R} \right\}$  \\
{\tt 5c:	receive(y);}\\
$\left\{(x_c, x_p, y) \in {\cal G}_{R} \right\}$  \\
\end{tabular}
\vspace{3mm}

We then encounter line {\tt 6c:  while (1)} which, together with line {\tt 12c: end} defines the main loop of the controller. We postulate the following pre- and post-conditions for these instructions

{\tt
\vspace{3mm}
\begin{tabular}{l}
$\left\{(x_c, x_p, y) \in {\cal G}_{R} \right\}$  \\
6c:  while (1)  \\
$\left\{(x_c, x_p, y) \in {\cal G}_{R} \right\}$  \\
\vdots \\
$\left\{(x_c, x_p, y) \in {\cal G}_{R} \right\}$  \\
12c:  end \\
$\left\{ false \right\}$.
\end{tabular}
\vspace{3mm}
}

The next line {\tt 7c: yc = max(min(y,1),-1)}, applies the nonlinear saturation operator to $y$. 
Common systems practice (as well as the methods used to compute $P$ in the first place) 
suggests that this nonlinear relation may be accurately captured by the quadratic inequality 
\[
(y_c-y)(y_c-0.2y) \leq 0,
\]
also named a {\em sector bound} on the saturation operator. This sector bound is not true in general, 
but it holds for all variables in ${\cal G}_R$. This bound may also be expressed in matrix form

\begin{equation}
\left[\begin{array}{c} x_c \\ x_p \\ y \\ y_c \end{array} \right]^T
\left[\begin{array}{cccc}
0 & 0 & 0 & 0\\
0 & 0 & 0 & 0 \\
0 & 0 & 0.2 & -0.6 \\
0 & 0 & -0.6 & 1
\end{array} \right]\left[\begin{array}{c} x_c \\ x_p \\ y \\ y_c \end{array} \right] = \left[\begin{array}{c} x_c \\ x_p \\ y \\ y_c \end{array} \right]^T
T\left[\begin{array}{c} x_c \\ x_p \\ y \\ y_c \end{array} \right]\leq 0.
\label{T-def}
\end{equation}

Thus one possible set of pre- and post- conditions for line 7c is 

{\tt
\vspace{3mm}
\begin{tabular}{l}
$\left\{(x_c, x_p, y) \in {\cal G}_{R} \right\}$  \\
{\tt 7c: yc = max(min(y,1),-1)}\\
$\left\{(x_c, x_p, y) \in {\cal G}_{R}, \; \left[\begin{array}{c} x_c\\x_p\\y\\y_c \end{array} \right]^TT
\left[\begin{array}{c} x_c\\x_p\\y\\y_c \end{array} \right] \leq 0\right\}$
\end{tabular}
\vspace{3mm}
}

However, following the principles outlined in~\cite{Fer1:08}, we seek to capture all variables within a single quadratic constraint, for the purpose of making proof checking simple and mirroring the usage 
of the ${\cal S}$-procedure to establish stability proofs at the specification level. To do 
so, we rely on the following lemma:

{\bf Lemma:} Assume the real vector variables $z$ and $w$ satisfy the constraints 
\[
\left[\begin{array}{cc}1 & z^T \\ z & U \end{array}\right] \geq 0
\]
for a given $U= U^T$ and
\[
\left[\begin{array}{c}z \\ w \end{array} \right]^TT\left[\begin{array}{c}z \\ w \end{array} \right]\leq 0
\]
for $T = T^T$. Then for any real $\mu$, we have 
\[
\left[\begin{array}{c} z \\ w \end{array} \right] \in {\cal G}_V ,\;\; V = \left(\left[\begin{array}{cc}I & 0 \\0 & 0\end{array}\right]-\mu \left[\begin{array}{cc} U & 0 \\0 & I\end{array} \right] T\right)^{-1}\left[\begin{array}{cc} U & 0 \\0 & I\end{array} \right]
\]
whenever $V$, defined as such, exists and is positive definite. 

{\bf Proof:} The proof is simple and therefore omitted.

Applying this lemma with $U=R$, $T$ defined as in~(\ref{T-def}) and $\mu = -\lambda = -6.76$
yields the inequality
\[
\left[\begin{array}{cc}
1 & [x_c^T \; x_p^T \; y \; y_c] \\
\left[\begin{array}{c} x_c \\ x_p \\ y \\ y_c \end{array} \right] & V 
\end{array} \right]\geq 0.
\]
and the pre- and post-conditions

\vspace{3mm}
\begin{tabular}{l}
$\left\{(x_c,x_p,y) \in {\cal G}_{R}\right\}$\\
{\tt 7c: yc = max(min(y,1),-1);}\\
$\left\{(x_c,x_p,y,y_c) \in {\cal G}_{V}\right\}$
\end{tabular}
\vspace{3mm} 

At this point, the variable $y$ is not of use anymore and we may therefore
release it. Thus we replace the post-condition $\left\{(x_c,x_p,y,y_c) \in {\cal G}_{V}\right\}$ of line 7c by the post condition
$\left\{(x_c,x_p,y_c) \in {\cal G}_{W}\right\}$, with 
\[
W = \left[\begin{array}{cccc} I & 0 & 0 & 0 \\
															0 & I & 0 & 0 \\
															0 & 0 & 0 & I \end{array}\right]V
															\left[\begin{array}{cccc} I & 0 & 0 & 0 \\																													0 & I & 0 & 0 \\
															0 & 0 & 0 & I \end{array}\right]^T
\]


The assignment {\tt 8c: u = Cc*xc + Dc*yc} is then easily commented as

\vspace{3mm}
\begin{tabular}{l}
$\left\{(x_c,x_p,y_c) \in {\cal G}_{W}\right\}$\\
{\tt 8c: u = Cc*xc + Dc*yc}\\
$\left\{(x_c,x_p,u,y_c) \in {\cal G}_{X}\right\}$
\end{tabular}
\vspace{3mm} 

with 
\[
X = \left[\begin{array}{ccc}
I & 0 & 0\\
0 & I & 0\\
C_c & 0 & D_c\\
0 & 0 & 1\end{array} \right]W\left[\begin{array}{ccc}
I & 0 & 0\\
0 & I & 0\\
C_c & 0 & D_c\\
0 & 0 & 1\end{array} \right]^T.
\]

The next line of the controller loop {\tt 9c: xc = Ac*xc + Bc*yc} can be treated similarly to line 8c, to yield the triple

\vspace{3mm}
\begin{tabular}{l}
$\left\{(x_c,x_p,u,y_c) \in {\cal G}_{X}\right\}$\\
{\tt 9c: xc = Ac*xc + Bc*yc}\\
$\left\{(x_c,x_p,u) \in {\cal G}_{Y}\right\}$
\end{tabular}
\vspace{3mm}

with 
\[
Y = \left[\begin{array}{cccc} Ac & 0 & 0 & Bc \\
 0 & I & 0 & 0\\
 0 & 0 & 1 & 0\\
 \end{array} \right]X \left[\begin{array}{cccc} Ac & 0 & 0 & Bc \\
 0 & I & 0 & 0\\
 0 & 0 & 1 & 0
 \end{array} \right]^T.
 \]
Note that the variable $y_c$ is not used anymore and has therefore been released. Line {\tt 10c} does not influence the variables and therefore can be commented as

\vspace{3mm}
\begin{tabular}{l}
$\left\{(x_c,x_p,u) \in {\cal G}_{Y}\right\}$\\
{\tt 10c: send(u)}\\
$\left\{(x_c,x_p,u) \in {\cal G}_{Y}\right\}$
\end{tabular}
\vspace{3mm}

The only remaining line in the controller loop is then line {\tt 11c: receive(y)}, which blocks the controller loop until a new value of $y$ is available. 

At that point, it becomes important to look again at the plant process, which restarts at line 
{\tt  7p:	receive(u)}. A necessary post-condition for this line is
$\left\{(x_c,x_p,u) \in {\cal G}_{Y}\right\}$, but no pre-condition can be clearly extracted. 
Line 7p will therefore be only partially instrumented, replacing the pre-condition by the symbol $\vdots$.
In addition, we indicate between brackets which line execution in the controller code (line 10c) unlocks 
line 7p, yielding the ``triple''

\vspace{3mm}
\begin{tabular}{l}
$\vdots $ \\
{\tt  7p:	  receive(u) [10c]}  \\
$\left\{(x_c,x_p,u) \in {\cal G}_{Y}\right\}$.
\end{tabular}
\vspace{3mm}

Consider then line {\tt  8p:  $\;$xp = Ap*xp + Bp*u;}. This assignment is properly documented with pre- and post- conditions to yield

\vspace{3mm}
\begin{tabular}{l}
$\left\{(x_c,x_p,u) \in {\cal G}_{Y}\right\}$\\
{\tt  8p:  $\;$xp = Ap*xp + Bp*u;}\\
$\left\{(x_c,x_p) \in {\cal G}_{Z}\right\}$
\end{tabular}
\vspace{3mm}

where 
\[
Z = \left[\begin{array}{ccc}
I & 0 & 0 \\
0 & A_p & B_p\end{array} \right]Y\left[\begin{array}{ccc}
I & 0 & 0 \\
0 & A_p & B_p\end{array} \right]^T,
\]
and $u$ has been released.

At that point, the plant dynamics meets line {\tt 9p: end}. That line has already been instrumented with candidate pre-and post conditions, which therefore must be shown to be compatible with the post-condition of line 9p. More precisely, we must show
\begin{equation}
(x_c, x_p) \in {\cal G}_Z \Rightarrow (x_c, x_p) \in {\cal E}_P.
\label{invariance}
\end{equation}
Numerical computations show that this is indeed true (modulo floating point operation arithmetic errors). 
While the entire loop describing the plant dynamics has been investigated, such is not the case of the controller dynamics, whose lines 11c and 12c have not been executed yet. We therefore need to keep tracking the execution of the two programs. While the control program is still blocked at line 11c, the plant dynamics loop sends the program execution to line 4p, 5p, and 6p, where the pre- and post- conditions are already established and coherent. Line 7p then blocks any further propagation of the plant dynamics, while the controller program is now allowed to proceed past line 11c, for which a valid post-condition is therefore $\{(x_c, x_p, y) \in {\cal G}_R \}$ and we will instrument as follows

\begin{tabular}{l}
\vdots \\
{\tt 11c: receive(y) [6p]} \\
$\left\{(x_c,x_p,y) \in {\cal G}_{R}\right\}$
\end{tabular}
\vspace{3mm}

which is compatible with the following line, 12c, which has been already documented.

The resulting commented programs then look like those shown in Table~\ref{complete_table}.

\begin{table}
{\tt \hspace{-20mm}
\begin{tabular}{l}
\% Controller Dynamics\\
$\left\{(x_c, x_p, y) \in {\cal G}_{R} \right\}$  \\
1c:  Ac = [0.4990, -0.0500; 0.0100,1.0000]; \\
$\left\{(x_c, x_p, y) \in {\cal G}_R \right\}$  \\
2c:  Cc=[564.48, 0];\\
$\left\{(x_c, x_p, y) \in {\cal G}_{R} \right\}$  \\
3c:  Bc=[1;0]; D = -1280;  \\
$\left\{(x_c, x_p, y) \in {\cal G}_{R} \right\}$  \\
4c:  xc = zeros(2,1); \\
$\left\{(x_c, x_p, y) \in {\cal G}_{R} \right\}$  \\
5c:  receive(y)  \\
$\left\{(x_c, x_p, y) \in {\cal G}_{R} \right\}$  \\
6c: while (1)\\
$\left\{(x_c, x_p, y) \in {\cal G}_{R} \right\}$  \\
{\tt 7c: yc = max(min(y,1),-1)}\\
$\left\{(x_c,x_p,y,y_c) \in {\cal G}_{V}\right\}$\\
\tt skip\\
$\left\{(x_c,x_p,y_c) \in {\cal G}_{W}\right\}$\\
{\tt 8c: u = Cc*xc + Dc*yc}\\
$\left\{(x_c,x_p,u,y_c) \in {\cal G}_{X}\right\}$\\
{\tt 9c: xc = Ac*xc + Bc*yc}\\
$\left\{(x_c,x_p,u) \in {\cal G}_{Y}\right\}$\\
10c: send(u)\\
$\left\{(x_c,x_p,u) \in {\cal G}_{Y}\right\}$\\
$\vdots$ \\
11c: receive(y) [6p] \\
$\left\{(x_c, x_p, y) \in {\cal G}_{R} \right\}$  \\
12c:  end \\
$\left\{ false \right\}$.
\end{tabular}
\begin{tabular}{l}
\% Plant Dynamics \\
$\left\{ (x_c, x_p) \in  {\cal E}_{P}\right\}$\\
1p:  Ap = [1.0000,0.0100;-0.0100,1.0000];\\
$\left\{ (x_c, x_p) \in  {\cal E}_{P} \right\}$\\
2p:  Cp=[1,0];\\
$\left\{ (x_c, x_p) \in  {\cal E}_{P} \right\}$\\
3p:  Bp = [0.00005; 0.01]; \\
$\left\{ (x_c, x_p) \in  {\cal E}_{P} \right\}$ \\
{\tt 4p:   while (1)}\\
$\left\{ (x_c, x_p) \in  {\cal E}_{P} \right\}$ \\
{\tt 5p:  y = Cp*xp}\\
$\left\{ (x_c, x_p, y) \in  {\cal G}_{R} \right\}$ \\
{\tt 6p: send(y);}\\
$\left\{(x_c, x_p, y) \in {\cal G}_{R} \right\}$  \\
$\vdots$\\
 7p: receive(u); [10c]\\
 $\left\{(x_c,x_p,u) \in {\cal G}_{Y}\right\}$\\
{\tt  8p:  $\;$xp = Ap*xp+ Bp*u;}\\
$\left\{(x_c,x_p) \in {\cal G}_{Z}\right\}$\\
\tt skip \\
$\left\{ (x_c, x_p) \in  {\cal E}_{P} \right\}$  \\
{\tt 9p: end}\\
$\left\{ false \right\}$
\end{tabular}
}
\caption{Commented closed-loop control program}
\label{complete_table}
\end{table}

\section{Sector-bounded nonlinearities and stability analysis}
Following similar developments in~\cite{Fer1:08}, there are apparent differences between the stability analysis performed in the requirements-level stability analysis, summarized by Eq.~(\ref{S-proc}), and the forward analysis performed afterwards. Namely, we want to show that the invariance condition~(\ref{invariance}) is satisfied if and only if the condition~(\ref{S-proc}) is satisfied. For that purpose, we begin with the forward analysis of the system, beginning at line {\tt 7c: yc = max(min(y,1),-1)}. The assertion
\[
\left\{(x_c, x_p, y) \in {\cal G}_{R}, \; \left[\begin{array}{c} x_c\\x_p\\y\\y_c \end{array} \right]^TT
\left[\begin{array}{c} x_c\\x_p\\y\\y_c \end{array} \right] \leq 0\right\}\]
with 
\[
{R} = 
\left[\begin{array}{cc} {P}^{-1} & {P}^{-1}C^T \\
C {P}^{-1} & C{P}^{-1}C^T 
\end{array} \right], \mbox{ and }
T=\left[\begin{array}{cccc}
0 & 0 & 0 & 0\\
0 & 0 & 0 & 0 \\
0 & 0 & 0.2 & -0.6 \\
0 & 0 & -0.6 & 1
\end{array} \right]
\]

implies the assertion 
\[
\left[\begin{array}{cc}
1 & [x_c^T \; x_p^T \; y \; y_c] \\
\left[\begin{array}{c} x_c \\ x_p \\ y \\ y_c \end{array} \right] & V 
\end{array} \right]\geq 0,
\]
with 
\[
V = \left(\left[\begin{array}{cc}I & 0 \\0 & 0\end{array}\right]-\mu \left[\begin{array}{cc} R & 0 \\0 & I\end{array} \right] T\right)^{-1}\left[\begin{array}{cc} R & 0 \\0 & I\end{array} \right].
\]

Long and somewhat tedious computations indicate that $W$, where $W$ is obtained by erasing the row and column of $V$ corresponding to the the variable $y$, satisfies
\[
\begin{array}{l}
W = \left[\begin{array}{cc} P+ 0.2 \mu C^TC & -0.6 \mu C^T \\
-0.6 \mu C & -\mu \end{array} \right]^{-1} \\[10pt]
= \left[\begin{array}{cc} \tilde{P}^{-1} & 0.6\tilde{P}^{-1}C^T \\ 0.6C\tilde{P}^{-1} & \displaystyle -\frac{1}{\mu} + 0.36\mu C\tilde{P}^{-1}C^T \end{array} \right]\\[10pt]
= \left[\begin{array}{cc} I & 0\\0.6C & 1\end{array} \right]\left[\begin{array}{cc}\tilde{P}^{-1} & 0 \\0 & -\displaystyle\frac{1}{\mu}\end{array} \right]\left[\begin{array}{cc}I & 0.6C^T \\ 0 & 1\end{array} \right],
\end{array}
\]
with $\tilde{P} = P-0.16 \mu C^TC$. Following the forward analysis and substituting matrix expressions for their values eventually yield
\[
Z = \left[\begin{array}{cc} A & B \end{array}\right]\left[\begin{array}{cc} I & 0\\0.6C & 1\end{array} \right]\left[\begin{array}{cc}\tilde{P}^{-1} & 0 \\0 & -\displaystyle\frac{1}{\mu}\end{array} \right]\left[\begin{array}{cc}I & 0.6C^T \\ 0 & 1\end{array} \right]\left[\begin{array}{cc} A & B \end{array}\right]^T.
\]
The invariance condition~(\ref{invariance}) therefore becomes
\[
\left[\begin{array}{cc} A & B \end{array}\right]\left[\begin{array}{cc} I & 0\\0.6C & 1\end{array} \right]\left[\begin{array}{cc}\tilde{P}^{-1} & 0 \\0 & -\displaystyle\frac{1}{\mu}\end{array} \right]\left[\begin{array}{cc}I & 0.6C^T \\ 0 & 1\end{array} \right]\left[\begin{array}{cc} A & B \end{array}\right]^T\leq P^{-1}
\]
Using Schur complements a first time yields the inequality
\[
\left[
\begin{array}{c|c}
-P^{-1} & \left[\begin{array}{cc}A&B\end{array} \right]\left[\begin{array}{cc} I & 0 \\0.6C & 1\end{array} \right] \\\hline [30pt]
\left[\begin{array}{cc} I & 0 \\0.6C & 1\end{array} \right]^T\left[\begin{array}{cc}A&B\end{array} \right]^T & \begin{array}{cc} -\tilde{P} & 0 \\ 0 & \mu \end{array} \end{array}\right]\leq 0.
\]
Using Schur complements a second time then yields the inequality
\[
\left[\begin{array}{cc} I & 0 \\0.6C & 1\end{array} \right]^T\left[\begin{array}{cc}A&B\end{array} \right]^TP\left[\begin{array}{cc}A&B\end{array} \right]\left[\begin{array}{cc} I & 0 \\0.6C & 1\end{array} \right]- \left[\begin{array}{cc} \tilde{P} & 0 \\ 0 & -\mu \end{array}\right]\leq 0.
\]
Multiplying this inequality to the left and right by
\[
\left[\begin{array}{cc} I & 0 \\-0.6C & 1\end{array} \right]^T \mbox{ and } \left[\begin{array}{cc} I & 0 \\-0.6C & 1\end{array} \right],
\]
respectively, yields 
\[
\left[\begin{array}{cc}A&B\end{array} \right]^TP\left[\begin{array}{cc}A&B\end{array} \right]-\left[\begin{array}{cc}P & 0\\0 & 0\end{array} \right]+ \mu \left[\begin{array}{cc} 0.2C^TC & -0.6C^T \\ -0.6C & 1 \end{array} \right] \leq 0,
\]
which is the same as~(\ref{S-proc}), with $\mu = -\lambda$.

\section{Conclusions}
This report describes an approach to documenting control programs, whereby the control program code is annotated with logical expressions describing the set of reachable program states. This approach constitutes the application of the Floyd-Hoare paradigm to control programs. It is shown that the the extensive domain knowledge gathered by control theory about control system specifications is readily applicable to develop stability and performance proofs of the corresponding control programs. Some system elements, such as sector-bounded nonlinearities, can be handled via forward analysis, and this analysis was shown to be equivalent to well known results in absolute stability theory. 

The analyses discussed in this report may be used in several different contexts: First, they may be used in an autocoding environment, whereby diagram-based specification languages such as Simulink can be autocoded in a target language such as C to automatically produce software with extensive, inlined proofs of software stability and performance. Such a documented software may then be easily verified independently by a proof checker.

Second, the analyses described in this paper may also be used to produce stability and poerformance proofs of undocumented software, even if little or no information is available about the software specifications. Producing such proofs becomes an integral part of the software verification process. Although time-consuming, this process is the only available option for much of the existing body of real-time control software. A particular case of interest is concerned with autocoded sofftware, where prior understanding of the autocoding process may help speed up the verification task.

\section*{Acknowledgements}

This report was prepared under partial support from the National Science Foundation under grant CSR/EHS 0615025, NASA under cooperative agreement NNX08AE37A, and the Dutton-Ducoffe professorship at Georgia Tech.


\newcommand{\etalchar}[1]{$^{#1}$}

\end{document}